\documentclass[12pt]{article}

\usepackage{amsmath,amssymb}

\oddsidemargin  - 5 mm
\evensidemargin - 5 mm

\newcommand{\be}{\begin{equation}}
\newcommand{\ee}{\end{equation}}
\newcommand{\bea}{\begin{eqnarray}}
\newcommand{\eea}{\end{eqnarray}}

\newcommand{\nn}{\nonumber}

\newcommand{\sot}{SO(3, \mathbb{R})}

\begin{document}

\title{\bf
Bianchi type I cosmology and the Euler-Calogero-Sutherland model}

\author{
A.M. Khvedelidze \\[0.3cm]  
{\it
Laboratory of Information Technologies},\\
\it{Joint Institute for Nuclear Research, 141980 Dubna, Russia}\\
{\it and
A. Razmadze Mathematical Institute, GE-380093 Tbilisi, Georgia}\\[0.5cm]
D.M. Mladenov\\[0.3cm]
{\it
Bogoliubov Laboratory of Theoretical Physics},\\
{\it Joint Institute for Nuclear Research, 141980 Dubna, Russia}
}

\date{(Received 1 September 2002)}
\maketitle

\begin{abstract}
{\small The Bianchi type I cosmological model is brought into a form  where
the evolution of observables is governed by the unconstrained Hamiltonian
that coincides with the Hamiltonian describing the relative motion of particles
in the integrable three-body hyperbolic Euler-Calogero-Sutherland system}.\\[0.3cm]
PACS number(s): 98.80.Hw, 04.20.Jb, 04.60.Ds
\end{abstract}


\section{Introduction}


Dealing with cosmological studies, the conceptual problem of the
identification of true degrees of freedom of gravity,
the ``golden fleece'' of canonical gravity \cite{IsenbergNester},
transforms into a real practical problem.
In the framework of the conventional formulation of Hamiltonian dynamics
of the cosmological models (see, e.g., \cite{Ryan,MTW,RyanShepley,Bogoyavlensky}),
based on the Dirac \cite{Dirac:1958jc} and the Arnowitt-Deser-Misner (ADM)
treatment of general relativity \cite{ADM},
the solution of the problem involves two ingredients:
choosing a certain coordinate fixing condition (gauge) and solving the constraints.
In the application to the minisuperspace models of Bianchi type,
the standard choice of the intrinsic time gauge leads to a reduced
Hamiltonian which is in general time-dependent and represents a square-root
expression \cite{Misner:1969ae,Ryan,MTW,RyanShepley}.
In spite of the fact that a ``relativistic'' physicist takes delight in this
square-root expression as a reminder of the relativistic particle energy,
the obtained result is not quite satisfactory.
Particularly, passing to the quantum theory, it brings the unpleasant difficulties of
defining the quantized square-root Hamiltonian as an operator in Hilbert space.

In the present paper, we would like to draw attention to the
fact that for the simplest Bianchi type I cosmological model,
it is possible to use an alternative more attractive generator of evolution,
free from the above-mentioned difficulties.
Below it will be shown that one can arrive at a reduced, quadratic in
momenta, time-independent Hamiltonian, in contrast to the above-mentioned
textbook result.
Moreover, a transparent correspondence between the resulting
deparametrized theory and a well-known nonrelativistic many-body integrable
system naturally arises.
Namely, it turns out that the expression for the generator of
evolution of the unconstrained Bianchi type I cosmology coincides with the
Hamiltonian describing the relative motion of the particles in the
three-body hyperbolic Euler-Calogero-Sutherland model.
\footnote{
This system is a generalization of the many-body Calogero-Sutherland model
\cite{Calogero1971,Sutherland1971,Sutherland1972,Moser}
(see also the useful reviews
\cite{OlshanetskyPerelomov1,OlshanetskyPerelomov2,PerelomovBook}),
describing particles on a line and interacting via pairwise hyperbolic potential
$1/\sinh^2 x$, by introduction of spin variables of the particles
\cite{GibbonsHermsen,Wojciechowski}.
For the other types of generalizations by inclusion of spinlike variables in
the Calogero-Sutherland model, see, e.g.,
\cite{PolychronakosDiscrete,Khvedelidze:2001mj,AKDM:PAN2002}.}

The plan of the paper is as follows.
The first part offers a short review of Bianchi cosmological models
including as well geometrical aspects as its Dirac generalized dynamics
\cite{DiracLectures,Sundermeyer,HenneauxTeitelboim,BorisovMamaev}.
Then, we restrict our consideration to the simplest Bianchi type I cosmology.
Analyzing the energy constraint and the extended Poincar\'{e}-Cartan 1-form,
we show how to deparametrize the theory, identify the physical degrees of freedom, and
as a result establish the correspondence with the three-body Euler-Calogero-Sutherland model.
Finally, using the relation between the dynamics of the
Euler-Calogero-Sutherland system and the geodesic motion on
the space of symmetric matrices, we discuss the general solution of the Bianchi type I model.


\section{Bianchi cosmology: geometry and dynamics}


The canonical formulation of the cosmological models
intensively uses the geometrical symmetry to restrict the
gravitational configuration space, the space of all
possible pseudo-Riemannian metrics defined on a given manifold.
In the case of large symmetry of the space-time manifold,
the gravitational degrees of freedom are effectively reduced to finite number
and this circumstance essentially relieves the analysis of the theory.
Below we describe very briefly this effective gravitational
configuration space for the homogeneous Bianchi type models.
For details, we refer to the comprehensive reviews
\cite{Ryan,RyanShepley,Bogoyavlensky}.


\subsection{Geometrical settings}


The conventional Hamiltonian analysis requires the existence of a
preferred timelike variable and therefore views the Universe in
terms of space plus time.
Below in quick review of the $3 + 1$ geometry of gravitation and its canonical formulation,
we follow the definitions and notations from the very transparent
review by Isenberg and Nester \cite{IsenbergNester}.

So, the space-time is supposed to be a smooth  manifold,
$({\cal M}\,, {\bf g})$ endowed with a metric ${\bf g}$ of signature $(-,+,+,+)$,
metric-compatible connection, and foliated by a family of one-parameter
three-dimensional nonintersecting surfaces $\Sigma_t$ on which
the induced metric has three positive eigenvalues, i.e.,
${\cal M}\,= \mathbb{R} \times \Sigma_t$.
According to such a space-time foliation, it is useful to choose on
\({\cal M}\) a surface-compatible frame of vector fields
\footnote{
Following \cite{IsenbergNester},
the boldface notation is used to distinguish four-dimensional quantities
from three-dimensional ones.}
$(\mbox{\bf e}_\bot \,, e_1,\, e_2, e_3)$
with timelike unit-length vector field $\mbox{\bf e}_\bot$
orthogonal to $\Sigma_t$ and three spacelike vector fields 
\footnote{
Hereafter, the latin indices run over 1,2,3.}
$e_a = (e_1, e_2, e_3)$tangent to it.
In the corresponding frame of forms 
$(\mbox{\boldmath $\theta$}^\bot \,, \theta^a)$, the metric \(\mbox{\bf g}\) reads
\begin{equation}  \label{eq:metricg}
\mbox{\bf g} = -
\mbox{\boldmath $\theta$}^\bot \otimes \mbox{\boldmath $\theta$}^\bot +
\gamma_{ab}\, \theta^a \otimes \theta^b\,,
\end{equation}
with a spatial metric \(\gamma \) induced on \(\Sigma_t\).
According to this representation, the Lie derivative
\({\cal L}_{\mbox{\bf e}_\bot} \), derivative with respect to
the proper time along the normal to the hypersurface \(\Sigma_t\), describes the
evolution of the dynamical variables.
Note that the frame $(\mbox{\bf e}_\bot , e_a)$ is not a coordinate frame and usually, 
instead of dealing with Eq. (\ref{eq:metricg}), 
the so-called ADM metric \cite{MTW} is exploited.
The latter is based on using the coordinate vector fields
$\mbox{\bf e}_0 = \mbox{\boldmath $\partial/\partial\, t$}$ and
$e_a = \partial / \partial \, x^a $, where the vector field \(\mbox{\bf e}_0 \)
is expressed in terms of the
normal vector field \( \mbox{\bf e}_\bot \) and spatial vector field
\( N^a \,e_a \) tangent to the hypersurface $\Sigma_t$ as
\begin{equation}
{\bf e}_0  = N\,{\bf e}_\bot + N^a\,e_a \,.
\end{equation}
However, when investigating the homogeneous Bianchi cosmological models, 
it is more convenient to deal with a special noncoordinate frame
instead of the ADM coordinate frame for the 3-space.
Its construction is dictated by geometrical considerations.
By definition, in a spatially homogeneous space-time, 
a three-dimensional Lie group \(G_3\) acts on the space-time as a group of isometries,
such that each orbit on which \(G_3\) acts simply transitively is a spacelike hypersurface.
The advantage of considering the simply transitive action is that we can
put the element of \(G_3\) into one-to-one correspondence with
the points of \(\Sigma_t \) and thus the space-time is considered
topologically as the product space ${\cal M} = {\mathbb{R}} \times G_3$.
Based on this observation, 
it is clear that instead of the usual coordinate frame of the spatial vector fields we need to
choose a new space basis \(e_a\), adapted to the Lie group structure of
the three-dimensional hypersurface \(\Sigma_t\).
Namely, the algebra of the infinitesimal generators of isometries,
i.e., the Killing vector fields $\xi_a$,
\begin{equation} \label{eq:LieAlgebra}
[\xi_a, \,\xi_b] =\, C^c_{\ ab}\, \xi_c\,,
\end{equation}
dictates what kind of basis should be chosen.
Indeed, the vector fields $\xi_a$  provide a basis
$(\mbox{\bf e}_0\,, e_a)$ invariant under the isometries
\begin{equation}
{\cal L}_{\xi_a}{\mbox{\bf e}_0} = 0\,, \qquad
{\cal L}_{\xi_a} {e_a} = 0\,.
\end{equation}
In this case, one can specify the form of the three-dimensional part of
the Bianchi metric as
\begin{equation}
\mbox{\boldmath $\gamma$} =
\gamma_{ab}\, {\omega}^a \otimes { \omega}^b
\,,
\end{equation}
where \({\omega}^a \) are group-invariant 1-forms dual to the vector fields $e_a$
with structure coefficients
$C^a_{\ bc} = 2\, d {\omega}^a (e_b, e_c)$,
defined by the structure constants of the homogeneity group
\(C^a_{\ bc}\).

Thus, instead of the ADM coordinate frame, 
the adapted representation for the Bianchi-type metrics in noncoordinate basis is
\begin{equation} \label{eq:BADM}
\mbox{\bf g} =
 - (N^2 - N^a\, N_a) \mbox{\bf dt} \otimes \mbox{\bf dt} +
2 N_a  \mbox{\bf dt} \otimes \omega^a + \gamma_{ab} \, {\omega}^a \otimes { \omega}^b\,.
\end{equation}
The preferable role of this choice for the frame is transparent:
from the Killing equation
\begin{equation}
{\cal L}_{\xi_a}
{\mbox{\bf g}}  = 0
\end{equation}
it follows that the functions $N, N^a$, and $\gamma_{ab}$,
entering in the expression for the Bianchi metric (\ref{eq:BADM}),
depend only on the time parameter $t$.


\subsection{Generalized dynamics}


As a result of this geometrical analysis, in an adapted basis, 
all gravitational field configuration variables, the lapse function \(N\), the
shift vector \( N^a \), and the spatial metric $\gamma_{ab}$ can be identified
with a set of ten Lagrangian coordinates of a certain  ``mechanical system.''

By definition, the dynamics of this system is determined from the Hilbert action
\begin{equation}\label{eq:actiong}
{\cal A } = \int_{\cal{M}} \mbox{\boldmath $\sigma$} \mbox{\bf R}\,,
\end{equation}
where the ansatz (\ref{eq:BADM}) is plugged into the
expression for the space-time scalar curvature \(\mbox{\bf R} \) and
for the four-dimensional volume element
\(
\mbox{\boldmath $\sigma $} =
\sqrt{-\mbox{\boldmath $g$}}\,
\,\omega^0\wedge\omega^1\wedge\omega^2\wedge\omega^3\,.
\)
As a result the restricted variational problem (for Bianchi class A models)
\footnote{
Bianchi class A models are defined writing the structure constants
of the isometry Lie group as
$C^d_{ab} = \epsilon_{lab} S^{ld} + A_{[ d} \delta^d_{b]}$
and supposing that $C^d_{ad} = A_a = 0$.
Here $S^{ab}$ is a three-dimensional second-rank symmetric tensor and
$A_a$ is a three-dimensional vector.
}
\cite{Ryan,RyanShepley}
is
\begin{equation} \label{eq:lagBian}
{\cal A }[N,N_a,\gamma_{ab},\dot{\gamma}_{ab}] =
\int\limits_{t_1}^{t_2}dt\sqrt{\gamma} N
\Bigl( {}^3\!R-K_a^{\ a}K_b^{\ b}+K_{ab}K^{ab}\Bigr) +
2\int\limits_{t_1}^{t_2}dt\ \frac{d}{d t}
\Bigl(\sqrt{\gamma} K^a_{\ a} \Bigr ),
\end{equation}
where \( {}^3\!R \) is the scalar curvature formed from the spatial metric \(\gamma \),
\begin{equation}
{}^3\!R= - \frac{1}{2}\,\gamma^{ab}C^c_{\ da}C^d_{\ cb} -
\frac{1}{4}\,\gamma^{ab}\gamma^{cd}\gamma_{ij} C^i_{\ ac}\,C^j_{\ bd},
\end{equation}
and
\begin{equation}
K_{ab}= - \frac{1}{2N}\,
\left[\;
(\gamma_{ad}\,C^d_{\ bc} + \gamma_{bd}\, C^d_{\ ac})\, N^c+
{\dot{\gamma}}_{ab} \;
\right]
\end{equation}
is the extrinsic curvature of the slice \(\Sigma_t\) defined as
\( K_{ab}= -\frac{1}{2}\,{\cal L}_{e_\bot}\gamma_{ab} \).

The Lagrangian (\ref{eq:lagBian}) belongs to the class of so-called
degenerate ones; its Hessian is zero.
To deal with its Hamiltonian description,
we shall follow the Dirac generalization of Hamiltonian dynamics
\cite{DiracLectures,Sundermeyer,HenneauxTeitelboim,BorisovMamaev}.

Implementing the Legendre transformation on variables
\(N,  N^a\), and  \({\gamma}_{ab}\), we get the canonical Hamiltonian
\begin{equation}
H_C = N {\cal H} + N^a{\cal H}_a 
\end{equation}
the primary
\begin{equation} \label{eq:constrp}
P = 0\,, \qquad   P_a = 0
\end{equation}
and secondary constraints
\begin{eqnarray}  \label{eq:constr}
&& {\cal H} = \frac{1}{\sqrt{\gamma}}
\left[\;\pi^{ab}\pi_{ab}-\frac{1}{2}\pi^a_{\ a}\pi^b_{\ b} \;\right]
- \sqrt{\gamma}\ {}^3\!R = 0\, ,  \\ \label{eq:mc}
&& { \cal H}_a = 2\ C^d_{\ ab}\pi^{bc}\gamma_{cd} = 0\,.
\end{eqnarray}
Here the canonical variables $(N\,, P), ( N^a\,, P_a),$ and $(\gamma_{ab}, \pi^{ab})$
obey the following nonvanishing fundamental Poisson brackets relations:
\begin{eqnarray}
&& \{N \,, P\} = 1\,,\\
&& \{N^a\,,P_b\} = \delta^a_b\,,\\
&& \{\gamma_{ab}\,, \pi^{cd}\}
= \frac{1}{2}\, \Bigl(\delta^c_a\, \delta^d_b + \delta^d_a\, \delta^c_b \Bigl)\,.
\end{eqnarray}
Due to the reparametrization symmetry of Eq. (\ref{eq:lagBian}) inherited from
the diffeomorphism invariance of the initial Hilbert action,
the evolution of the system is unambiguous and it is governed
by the total Hamiltonian
\begin{equation} \label{eq:totham}
H_T = N {\cal H} + N^a{\cal H}_a  + u P + u^a P_a\,,
\end{equation}
with four arbitrary functions \( u (t) \) and \( u^a (t)\).
One can verify that the secondary constraints are first class and
obey the algebra
\begin{eqnarray}  \label{eq:nonabalg}
&&\{{\cal H}\,, {\cal H}_b\} = 0, \\
&&\{{\cal H}_a\,, {\cal H}_b\} = - C^d_{\ ab}{\cal H}_d\,.
\end{eqnarray}
Thus the dynamics of system (\ref{eq:lagBian}) in the extended phase
space, spanned by the canonical variables $(N\,, P), (N^a\,, P_a)$, and
$(\gamma_{ab}\,,\pi^{ab})$,
is described by the action rewritten in the Hamiltonian form as
\begin{equation}  \label{eq:PCEXT}
{\cal A}  = \int  \Theta \  + \int d (\pi^{ab}\gamma_{ab})\,,
\end{equation}
where $\Theta$ is the Poincar\'{e}-Cartan 1-form
\begin{equation}
\label{eq:fPC}
\Theta =  \pi^{ab}\, d \gamma_{ab} + P\, dN + P_a\, dN^a - H_T\, dt\,.
\end{equation}


\section{Deparametrized version of the model}



\subsection{Misner-Ryan decomposition}


Let us now specialize to the Bianchi type I model.
In this case, the three-dimensional group of isometries is
an Abelian one, i.e., $C^d_{\ ab} = 0$, and thus the momentum constraints
(\ref{eq:mc}) are identically satisfied, while the energy constraint reduces
to the simplest form
\begin{equation}  \label{eq:enconI}
{\cal H} = \pi^{ab}\pi_{ab} - \frac{1}{2}\pi^a_{\ a}\pi^b_{\ b} \,.
\end{equation}

To find the corresponding unconstrained Hamiltonian system
with a certain observable time parameter, 
we shall use the variables introduced by Misner \cite{Misner:1969ae}
and modified by Ryan \cite{Ryan}.
In Misner's representation, the spatial metric $\gamma$ is given by
\begin{equation}\label{eq:MisnerRyan}
\gamma_{ij} = r_0^2 e^{-2\Omega} e^{2 \beta}_{\ \ ij}\,,
\end{equation}
where \(\beta_{ij}\) is a \(3\times 3\) symmetric traceless matrix
and \(\Omega\) is a scalar, both being functions of the time parameter only,
and $r_0$ is a constant.
\footnote{
In Eq. (\ref{eq:MisnerRyan}), the constant $r_0$ was introduced for convenience in choosing units.
For notational convenience, in this paper the constant $r_0$ is scaled
to unity hereafter.}
For our purposes it is convenient, following \cite{Ryan,RyanShepley},
to pass to the main-axes decomposition for the nondegenerate symmetric matrix
\(\beta\) and finally write the metric $\gamma$ in the
Misner-Ryan form as
\begin{equation}\label{eq:mainax}
\gamma =
R^T(\chi)
\left(
\begin{array}{ccc}
e^{\beta_+  + \sqrt{3}\beta_-} &             0                 &    0       \\
              0                & e^{\beta_+ - \sqrt{3}\beta_-} &    0       \\
              0                &             0                 & e^{-2\beta_+}
\end{array} \right )
R(\chi)\,.
\end{equation}
Here $R(\chi) $ is an orthogonal $\sot$ matrix parametrized with the three  Euler angles $\chi_i$.
From the Jacobian of the transformation (\ref{eq:mainax}),
\begin{equation}
J\left(\frac{\gamma_{ab} (\Omega, \beta, \chi)}{\Omega, \beta, \chi} \right)
\propto \ e^{-6\Omega} \mid
\sinh( 2\sqrt{3}\beta_-)
\sinh(3\beta_+ +\sqrt3 \beta_-)
\sinh(3\beta_+ -\sqrt3 \beta_-)
\mid\,,
\end{equation}
it follows that Eq. (\ref{eq:mainax}) can be used as a definition of
the new configuration variables,
($\Omega, \beta_\pm)$, and the three angles $(\chi_1, \chi_2,\chi_3$)
only if all eigenvalues of the matrix $\gamma$ are different.
To have a uniqueness of the inverse transformation, 
we are forced to treat only this type of configurations, the so-called principle
orbits of the action of the $\sot$ group,
while the analysis of the orbits with coinciding eigenvalues of the matrix $S$
(singular orbits) requires special consideration.

The point transformation (\ref{eq:mainax}) induces the canonical
transformation from the coordinates $(\gamma_{ab}\,, \pi^{ab})$
to the six new canonical pairs $(\Omega\,, p_\Omega )$,
$(\beta_\pm \,, p_\pm)$, and $(\chi_i \,, p_{\chi_j})$,
\begin{eqnarray}
&& \{ \Omega\,, p_\Omega\} = 1\,, \qquad
\{ \beta_\pm \,, p_\pm \} = 1\,, \qquad
\{ \chi_i \,, p_{\chi_j} \} = \delta_{ij}\,.
\end{eqnarray}
Using the requirement of invariance of the symplectic 1-form,
\begin{equation}
\pi^{ab}\, d\gamma_{ab} = p_\Omega \, d\Omega + p_+\, d\beta_+ + p_- d\beta_- +
\sum_{i=1}^3 p_{\chi_i}\, d p_{\chi_i}\,,
\end{equation}
one can find the explicit formulas for the new variables as functions of the old
ones and finally rewrite the energy constraint as
\begin{eqnarray} \label{eq:hamcon}
&& 
2\, {\cal H}  =  \frac{1}{12}\, (-p^2_\Omega +p_+^2+p_{-}^2)\nn\\
&& 
+ \frac{1}{4} \left(
\frac{\eta_1^2}{\sinh^2( 3\beta_+ -\sqrt3 \beta_-)} +
\frac{\eta_2^2}{\sinh^2(3\beta_+ +\sqrt3 \beta_-) }
+ \frac{\eta_3^2}{\sinh^2(2 \sqrt 3 \beta_-)}
\right)\,.
\end{eqnarray}
Note that in this expression all angular variables
are gathered into the three right-invariant $\sot$ Killing vector fields $\eta_i$,
\begin{eqnarray}
&& \label{eq:rf1}
\eta_1 =
- \sin\chi_1 \cot\chi_2 \ p_{\chi_1} +
\cos\chi_1 \  p_{\chi_2} +
\frac{\sin\chi_1}{\sin\chi_2}\ p_{\chi_3} \,,\\
&& \label{eq:rf2}
\eta_2 =
\,\,\cos\chi_1 \cot\chi_2 \ p_{\chi_1} +
\sin\chi_1 \  p_{\chi_2} -
\frac{\cos\chi_1}{\sin\chi_2}\ p_{\chi_3} \,, \\
&& \label{eq:rf3}
\eta_3 =  p_{\chi_1}\,,
\end{eqnarray}
satisfying the $\sot$ Poisson brackets algebra
\begin{equation}
\{\eta_i,\  \eta_j\} = \epsilon_{ijk} \eta_k\,.
\end{equation}


\subsection{
Deparametrization and correspondence to the
hyperbolic three-particle Euler-Calogero-Sutherland model}


In this section, 
it will be shown that the unconstrained Hamiltonian,
describing the evolution of observables in Bianchi type I cosmology, 
can be identified with the Hamiltonian of the relative motion in the
three-particle hyperbolic Euler-Calogero-Sutherland system.

The direct method to ``deparametrize'' the initial reparametrization-invariant theory 
with first-class constraints consists of two steps, namely
fixation of the gauge and projection of the initial degenerate action
to the constraint shell.
Then the expression for the reduced, unconstrained Hamiltonian can be extracted
from the Poincar\'{e}-Cartan 1-form.
Here it is in order to make one general comment on the gauge-fixing
when some preferred set of variables is chosen.
It turns out that in this case the action projected on the constraint shell
becomes independent of the part of the variables
without imposing some gauge-fixing condition.
These eliminated variables should naturally be treated as unphysical ones.
Such a phenomenon is well-known from the theory of Hamiltonian systems
possessing a Lie group symmetry; 
when the cordinates adapted to the symmetry group action are found, 
then the so-called cyclic (or ignorable) coordinates disappear from the effective Hamiltonian.
In the reparametrization-invariant theories, there is a certain peculiarity of this effect.
Namely, when such a set of well-defined variables exists,
then after reduction, one of these cyclic variables takes the role of
time in the corresponding deparametrized system.
The existence of such an exceptional variable in the reparametrization-invariant 
theories means that the equivalent unconstrained system can
be written in an autonomous form.

Below we shall see that the model under consideration
can be brought into a class of such Hamiltonian systems.
\footnote{
The corresponding representation for the Friedmann cosmological
model with scalar and spinor fields has been discussed recently
in \cite{Khvedelidze:2001tr}.}
To see how the proper variable which takes the role of time in the unconstrained system appears,
let us consider the energy constraint (\ref{eq:hamcon})
in terms of the Misner-Ryan coordinates.
One can easily verify that performing the canonical transformation
\begin{equation}
T = 6\ \frac{\Omega}{P_\Omega} \,, \qquad \Pi_T = \frac{1}{12} P_\Omega^2\,,
\end{equation}
the energy constraint (\ref{eq:hamcon}) linearizes with respect to the momentum $\Pi_T$,
\footnote{
Such a linearity in one of the momenta forms of the constraints in accordance
with Kucha\v{r}'s observation \cite{Kuchar}
shows the possibility of a global deparametrization of the degenerate theory.}
\begin{equation} \label{eq:hamconn}
2\, {\cal H} = - \Pi_T + h(\beta_\pm\,, p_\pm\,,\eta_i)\,,
\end{equation}
where $h(\beta_\pm\,, p_\pm\,,\eta_i)$ is given by
\begin{eqnarray} \label{eq:BiPh}
&&
h (\beta_\pm\,, p_\pm\,,\eta_i)= \frac{1}{12}\, (p_+^2 + p_-^2)\nn\\
&&
+ \, \frac{1}{4}\, \left(
\frac{\eta_1^2}{\sinh^2( 3\beta_+ -\sqrt3 \beta_-)} +
\frac{\eta_2^2}{\sinh^2(3\beta_+ +\sqrt3 \beta_-) } +
\frac{\eta_3^2}{\sinh^2(2 \sqrt 3 \beta_-)}
\right)\,.
\end{eqnarray}
Note that after all of these transformations, the Poincar\'{e}-Cartan 1-form
(\ref{eq:fPC}) changes by a total differential,
\begin{equation} \label{eq:PCEN}
\Theta =
\Pi_T dT  + p_+\, d\beta_+ + p_-\, d\beta_- + \sum_{i=1}^3 p_{\chi_i} d{\chi_i} +
P dN + P_a dN^a - H_T d t - \frac{1}{2}\,d(P_\Omega\Omega)\,.
\end{equation}

Now to find the evolution parameter and the Hamiltonian of the deparametrized theory, 
we project Eq. (\ref{eq:PCEN}) to the constraint shell $P = P_a = {\cal H}={\cal H}_a = 0$,
\footnote{
We omit all differentials assuming that proper boundary
conditions are taken to derive the classical equation of motion.}
\begin{equation} \label{eq:pPC}
\Theta\ \bigl\vert_{constraints} \ =
p_+ d \beta_+ + p_- d\beta_-  + \sum_{i=1}^3 p_{\chi_i} d{\chi_i}  +
h(\beta_\pm\,, p_\pm\,,\eta_i) d T \,.
\end{equation}
The expression for the projected 1-form (\ref{eq:pPC})
prompts the following interpretation.
Supposing that the deparametrized version of the Bianchi type I model
has a conventional Poincar\'{e}-Cartan 1-form,
\begin{equation}
\Theta^\ast _{BI} = p_+ d \beta_+  +  p_- d\beta_-
+  \sum_{i=1}^3 p_{\chi_i} d{\chi_i}\   -  H_{BI}(\beta_\pm\,,
p_\pm\,,\eta_i) d \tau
\,,
\end{equation}
we make the following identifications:
the initial dynamical variable $T(t)$ ``metamorphosis''
to the evolution parameter $\tau = -T$ of the corresponding deparametrized
system and the Hamiltonian, as a time-independent generator of evolution,
is given by  $H_{BI}= h(\beta_\pm\,, p_\pm\,,\eta_i)$.
Thus to summarize, 
we arrive at the unconstrained Hamiltonian for the Bianchi type I model in the form
\begin{eqnarray} \label{eq:BIUH}
&& H_{BI} = \frac{1}{12}\, (p_+^2 + p_{-}^2) \nn\\
&& + \  \frac{1}{4}\, \left(
\frac{\eta_1^2}{\sinh^2( 3\beta_+ -\sqrt3 \beta_-)} +
\frac{\eta_2^2}{\sinh^2(3\beta_+ +\sqrt3 \beta_-) } +
\frac{\eta_3^2}{\sinh^2(2 \sqrt 3 \beta_-)}
\right)\,.
\end{eqnarray}

One final comment is in order concerning the form of the derived Hamiltonian.
As was mentioned in the Introduction, 
the appearance of the square-root expression for the unconstraned
Hamiltonian of the Bianchi cosmologies is
usually interpreted in a certain analogy to relativistic particles
\cite{Ryan,RyanShepley}.
In contrast to this picture, our analysis points to the
existence of a certain correspondence with a system of nonrelativistic particles.

To describe this correspondence, let us recall the
formulation of the integrable three-particle hyperbolic Euler-Calogero-Sutherland (ECS)
model \cite{GibbonsHermsen,Wojciechowski}.
The particles are characterized by two types of variables.
The first ones are the canonical coordinates  $(\beta_i\,, p_i)$ describing
the position and momenta,
\begin{equation}
\{\beta_i, p_j\} = \delta_{ij}\,,
\end{equation}
and the other are the so-called  ``internal'' coordinates $l_{ab} = - l_{ba}$,
obeying the $\sot$ Poisson brackets algebra,
\begin{equation}
\{l_{ab}, l_{cd} \} =
\delta_{ac} l_{bd} - \delta_{ad} l_{bc} +
\delta_{bd} l_{ac} - \delta_{bc} l_{ad}
\,.
\end{equation}
The dynamics of both variables is determined by the Hamiltonian
\begin{equation} \label{eq:ECS}
H_{ECS}  =
\frac{1}{2} \sum_{i=1} ^3 p_i^2 + \frac{1}{8} \sum_{(i \neq j)}
\frac{l_{ij}^2}
{\sinh^2(\beta_i - \beta_j)}\,.
\end{equation}

Now it is easy to find the relation between the derived unconstrained
Hamiltonian (\ref{eq:BIUH}) and the Euler-Calogero-Sutherland Hamiltonian
(\ref{eq:ECS}).
To achieve this, we identify the internal variables $l_{ij}$  with
the Killing vectors $\eta_i$ of the Bianchi type I model,
\begin{equation}
l_{ij} =\epsilon_{ijk}\eta_k\,,
\end{equation}
while the variables $(\beta_\pm , \ p_\pm )$ are identified
with the relative variables in the Jacobi system for $(\beta_i \,, p_i)$,
\begin{eqnarray}
&\beta_1 =  \beta_+ +\sqrt3 \beta_{-}- X\,,\;\;\;\;\;\;\;\;\;\;&
p_1 =\frac{1}{6}\ p_+ + \frac{1}{2\sqrt 3}\ p_{-}-\frac{1}{3}\  P\,,\\
&\beta_2 = \beta_+ -\sqrt 3 \beta_{-} - X\,,\;\;\;\;\; \;\;\;\;\; &
p_2 = \frac{1}{6}\ p_+ - \frac{1}{2\sqrt 3}\ p_{-}-\frac{1}{3} \ P\,,\\
& \beta_3 = - 2 \beta_+ -  X\,,\;\;\;\;\; \;\;\;\;\; \;\;\;\;\;\;\;\;\;&
p_3 =-\frac{1}{3}\ p_+ -\frac{1}{3}\ P\,.
\end{eqnarray}
Here $(X, P)$ are the corresponding center-of-mass coordinates.
Now one can verify that the expression (\ref{eq:ECS})
takes the following separable form:
\begin{equation} \label{eq:E_SJ}
H_{ECS} = \frac{1}{6}\, P^2 + H_{BI}\,
\end{equation}
and thus one arrives at the above-stated correspondence
between the Hamiltonian, describing the relative motion in a three-particle
hyperbolic Euler-Calogero-Sutherland system, and
the deparametrized version of the Bianchi type I model Hamiltonian $H_{BI}$.


\subsection{Solution of the unconstrained equations of motion}


The established relation to the many-body integrable model allows us to
write down the general solution for the Bianchi type I metric.
Here it is in order to note that the solution of the initial Einstein
equations for the Bianchi-type metrics contain arbitrary functions,
reflecting the reparametrization invariance,
of the theory and the solution should be classified according to their equivalence.
Passing to the reduced system, when the constraints generating this
reparametrization are eliminated, this freedom is fixed.
However, if  the system has the additional rigid symmetry,
not all of solutions describe the nonequivalent metrics and now
the problem of classification reduces to the identification of the metrics,
with a fixed number of parameters.
Below we shall present the general solution for the reduced Bianchi type I model,
while the detailed analysis of the essential parameters and
their origin will be done elsewhere.

To write down an explicit form of the solution
for the general  metric of Bianchi type I space-time, 
we recall the roots of the integrability of the Euler-Calogero-Sutherland model.
It is known
(see, e.g., \cite{Khvedelidze:2001mj,AKDM:PAN2002} and references therein)
that the integral curves of the Hamiltonian (\ref{eq:ECS}) are in one-to-one correspondence
with the geodesics of the space of symmetric matrices $S$
endowed with the bi-invariant metric,
\begin{equation}
\label{eq:ecs}
L_{ECS} = \frac{1}{2}\, \int \mbox{Tr}(S^{-1} \, dS)^2\,.
\end{equation}
To prove this dynamical equivalence, it is enough
to use again the main-axes decomposition for the symmetric matrix
$S$ of the form
\begin{equation}
\label{eq:nma}
S = R^T\, e ^{2X}\, R
\end{equation}
with the diagonal matrix $X = \mbox{diag}\|\beta_1, \beta_2, \beta_3 \|$.
Assuming that $S_{ij}$ are six independent Lagrange coordinates,
from the action (\ref{eq:ecs}) it follows that the equation of motion can be
written in the form of conservation law,
\begin{equation}
\frac{d}{dt}\left(S^{-1}\frac{d}{dt}\, S \right ) = 0\,,
\end{equation}
and therefore the general solution for the matrix $S$
can be represented in a simple form,
\begin{equation}\label{eq:sol}
S(t) = S_0\, e^{J t}\,.
\end{equation}
Here $S_0$  and $J$ are arbitrary constant matrices,
$S_0$ is a symmetric matrix,  the value of the matrix $S(t)$ at an initial moment $t=0$,
while $J$ is an integral of motion,
$J = S^{- 1}_0\dot{S}_0$.
\footnote{
One can easily check that the
solution (\ref{eq:sol}) is a symmetric matrix if its initial value
$S_0$ ware a symmetric one.}
Comparison of the two expressions (\ref{eq:nma}) and (\ref{eq:sol}) gives the
solution of the Euler-Calogero-Sutherland model and therefore of the Bianchi type I
metric due to the correspondence derived in the previous section.


\section{Concluding remarks}


Following the Dirac generalized Hamiltonian approach we derive
the deparametrized version of the Bianchi type I cosmology
in the form of the integrable three-particle Euler-Calogero-Sutherland model.
It is worth posing the question of the possibility to extend the
result presented here and to find the quadratic time-independent
unconstrained Hamiltonians for the other more complicated Bianchi systems.
In seeking the answer, first of all it is necessary
to investigate the rigid symmetries that Bianchi models possess,
because the additional rigid symmetry as a
rule leads to the conception of preferred time choice \cite{Kuchar}.
More precisely, the existence of a timelike Killing vector field
allows us to define an intrinsic time and construct the
observables, whose dynamics is governed by a time-independent Hamiltonian.
\footnote{
Note that sometimes it is enough to have even a conformal
timelike Killing vector field; see, e.g., \cite{Khvedelidze:2001tr}.}

Another application of the discussed correspondence is to analyze
the metric of Bianchi type I cosmology with respect
to its dependence on the essential parameters.
In a forthcoming publication, we plan to discuss in detail a certain
parametrization of the metric using the solution (\ref{eq:sol}) of
the three-body Euler-Calogero-Sutherland problem and we shall attempt to clarify how
the topology of the flat Bianchi type I cosmology emerges.
One can find the recent discussions of the topological characteristics
of the three-dimensional space in the cosmology in review
\cite{Lachieze-ReyLuminet}.


\section*{Acknowledgments}


Helpful discussions during the work on the paper with
H. Braden, R. Camassa, G. Dvali, P. Hajicek, G. Lavrelashvili, M.D. Mateev,
V.P. Pavlov, O. Ragnisco, S. Ruijsenaars, and  Y. Suris are acknowledged.
A.K. would like to thank the Abdus Salam International Centre for Theoretical
Physics (ICTP), Trieste, Italy for support and kind hospitality extended
to him at ICTP High Energy Section.
D.M. is grateful to the organizers, with special thanks to Orlando Ragnisco, of the workshop
"The Calogero-Moser System Thirty Years Later," Roma, and to the
participants for many interesting conversations and discussions.
This work was supported in part by the RFBR grant No. 01-01-00708
and INTAS grant No. 00-00561.


\end{document}